\documentclass[%
superscriptaddress,
preprint,
 amsmath,amssymb,
aip,apl,
showkeys,
]{revtex4-2}
\usepackage[dvipdfmx]{graphicx}
\usepackage{dcolumn}
\usepackage{bm}
\usepackage{color}
\usepackage{txfonts}

\begin{document}

\preprint{PRESAT-9601}

\title{Density functional theory study on effect of NO annealing for SiC(0001) surface with atomic-scale steps}

\author{Mitsuharu Uemoto}
\affiliation{Department of Electrical and Electronic Engineering, Graduate School of Engineering, Kobe University, Nada, Kobe 657-8501, Japan}
\author{Nahoto Funaki}
\affiliation{Department of Electrical and Electronic Engineering, Graduate School of Engineering, Kobe University, Nada, Kobe 657-8501, Japan}
\author{Kazuma Yokota}
\affiliation{Department of Electrical and Electronic Engineering, Graduate School of Engineering, Kobe University, Nada, Kobe 657-8501, Japan}
\author{Takuji Hosoi}
\affiliation{School of Engineering, Kwansei Gakuin University, Sanda, Hyogo 669-1330, Japan}
\author{Tomoya Ono}
\affiliation{Department of Electrical and Electronic Engineering, Graduate School of Engineering, Kobe University, Nada, Kobe 657-8501, Japan}

\date{\today}

\begin{abstract}
Density functional theory calculations for the electronic structures of the 4H-SiC(0001)/SiO$_2$ interface with atomic-scale steps are carried out to investigate the effect of NO annealing. The characteristic behavior of the conduction band edge states of SiC is strongly affected over a wide area of the interface by the Coulomb interaction of the O atoms in the SiO$_2$ region as well as the step structure of the interface, resulting in the discontinuity of the inversion layers at the step edges under the gate bias. The spatially discontinued band only allows the very limited conduction paths in the inversion layer, leading to the significantly decreased mobile carrier density. It is found that the Coulomb interaction of the O atoms is screened and the inversion layers become continuous when the nitrided layers are inserted at the interface by NO annealing. This result is in good agreement with experimental findings that the improvement of the performance of SiC metal-oxide-semiconductor field-effect-transistors by NO annealing is attributed to an increase in the mobile electron density rather than an increase in the mobility of electrons in the inversion layer.
\end{abstract}

\maketitle


SiC is a technologically important material for future power electronic devices with high breakdown field, high carrier velocity, and thermally oxidized SiO$_2$ layers, making it ideal for metal-oxide-semiconductor field-effect transistors (MOSFETs). Among numerous of polymorphs of SiC (e.g., 3C, 4H, and 6H), 4H-SiC is the most commonly used in actual devices, which can be grown as single-polymorph wafers.\cite{JpnJApplPhys_45_007565,JpnJApplPhys_54_040103,JpnJApplPhys_54_04DP07,ApplPhysExpress_13_120101} However, the high channel resistance of SiC-MOSFETs limits their performance.\cite{PhysStatusSolidiA_162_321,ApplPhysLett_77_003281} This high resistance is expected to be attributed to the low field-effect mobility in SiC-MOSFETs, which is much lower than the ideal electron mobility ($\sim$ 1000 cm$^2$V$^{-1}$s$^{-1}$).\cite{MaterSciForum_433-436_443} By interface nitridation by post-oxidation annealing with nitric oxide (NO),\cite{IEEEElecDevLett_22_000176,ApplPhysLett_70_002028} one can increase the maximum field-effect mobility from 1-7 cm$^2$V$^{-1}$s$^{-1}$\cite{JApplPhys_91_001568} to 25-40 cm$^2$V$^{-1}$s$^{-1}$ for 4H-SiC(0001) MOSFETs.\cite{IEEETransElectronDevices_60_001260, IEEETransElectronDevices_62_000309,ApplPhysExpress_10_046601} Hatakeyama {\it et al.} reported that the Hall mobility is not increased by NO annealing; on the other hand, the mobile carrier density increases and the increase in field-effect mobility is due to the increase in mobile carrier density.\cite{ApplPhysExpress_12_021003} It is also reported that the inserted N atoms become a source of interface defects\cite{JApplPhys_112_024520} and/or cause negative bias instability.\cite{MaterSciForum_858_599} Although a lot of efforts have been made thus far, the microscopic information of such an increase in field-effect mobility is not fully clear.

The behavior of the conduction band edge (CBE) states of a SiC bulk is similar to that of free electrons, which are the so-called ``floating states,''\cite{PhysRevLett_108_246404} and sensitive to the local atomic structure at the interface.\cite{JPhysSocJpn_85_024701} In practical devices, off-oriented 4H-SiC(0001) surfaces by 4 degree are widely used and thus intrinsically possesses atomic-scale step and terrace structures even for the atomically flat surfaces.\cite{SSDM_1987_227} Furthermore, it has been reported that thermal oxidation causes atomic-scale roughness even on the oriented 4H-SiC(0001) surfaces,\cite{JApplPhys_106_123506} suggesting that atomic steps also exist on the terrace at the SiO$_2$/SiC interface. Atomic-scale steps affect the behaviors of the electrons of the floating states as well as the electronic structures of the interface over a wide area of the interface. In this paper, we compare the local densities of states (LDOSs) of the SiC(0001)/SiO$_2$ interface with steps before and after NO annealing using first-principles calculations based on the density functional theory (DFT).\cite{PhysRev_136_B864} It is found that the CBE states are affected by the Coulomb interaction of the O atoms in the SiO$_2$ region, which results in the discontinuity of the inversion layers at the step edge under the gate bias. The discontinuities of the inversion layers markedly prevent the carriers from conducting from the source to the drain. After annealing, the inserted nitrided layers screen the Coulomb interaction and the density of the spatial discontinuities of the inversion layer decreases. These results suggest that the nitrided layers inserted by NO annealing make the inversion layers continuous by screening the Coulomb interaction of the O atoms in the SiO$_2$ regions so that the penetration of carriers is enhanced.

\begin{figure}[htbp]
\includegraphics{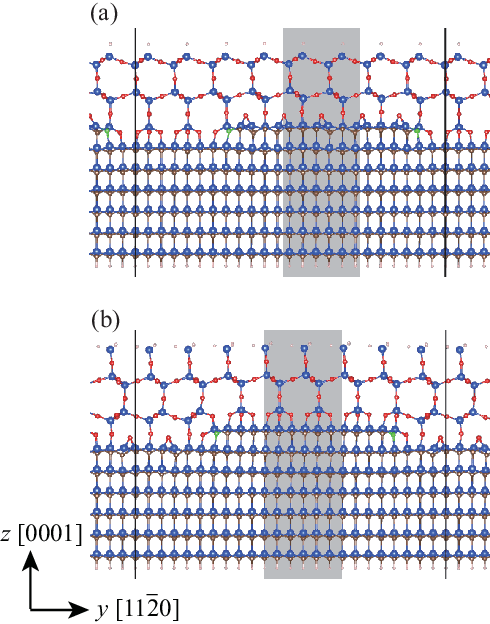}
\caption{Interface atomic structure with steps for (a) k1/h3 (h1/k3) and (b) k3/h1 (h3/k1) models. Green, blue, red, and white balls represent N, Si, C, and O atoms, respectively. Black lines are the boundaries of supercells. The LDOS in the shaded area is calculated. \label{fig:1}}
\end{figure}

We perform the first-principles calculation for the electronic structures of 4H-SiC(0001)/SiO$_2$ interfaces with steps before and after NO annealing. As an example, we show the computational models where the trench structure models are employed to imitate the step model as shown in Fig.~\ref{fig:1}. Since most of SiO$_2$ in the SiC(0001)/SiO$_2$ interface is amorphous as shown by a scanning transmission electron microscopy (STEM) image,\cite{PhysRevB_96_115311} it is not straightforward to characterize the interface atomic structure. Here, we assume the atomic structures that can exist locally at the SiC(0001)/SiO$_2$ interface and employ the interface atomic structures proposed in our previous study on the basis of STEM images.\cite{PhysRevB_96_115311}  Our model contains a crystalline substrate with seven SiC bilayers (17.5 \AA~thick) connected without any coordination defects to a crystalline SiO$_2$ region with a thickness of 8.2 \AA~for the upper terrace. In a 4H-SiC bulk, SiC bilayers are stacked on the quasi-cubic ($k$) and hexagonal ($h$) sites alternatively along the [0001] direction. For the SiO$_2$ side, the computational models where the 4H-SiC(0001) bilayer faces the one-hold or three-hold structure of SiO$_2$ are proposed in our previous study.\cite{PhysRevB_96_115311} The interface model, where the atoms in the $k$ site of the 4H-SiC(0001) bilayer face the one-hold SiO$_2$ structure, is referred to as the k1 model. The other interface models are also similarly named the k3, h1, and h3 models.

For the interfaces with steps, when the upper terrace of the 4H-SiC(0001) bilayer is at the $h$ site, the lower terrace is always at the $k$ site. In addition, when the SiO$_2$ layer at the upper terrace has the three-hold structure, that at the lower terrace has the one-hold structure. The dangling bonds at the step edges are removed by replacing C atoms with N atoms. The interface model with steps, in which the k1 and h3 interfaces are the upper and lower terraces, respectively, is named the k1/h3 model. The other interface models with steps are similarly referred to the k3/h1, h1/k3, and h3/k1 models.

The RSPACE code,\cite{PhysRevLett_82_005016,KikujiHirose2005,PhysRevB_82_205115} which uses the real-space finite-difference approach\cite{PhysRevLett_72_001240,PhysRevB_50_011355} for the DFT,\cite{PhysRev_136_B864} is employed. The periodic boundary condition is imposed on all the directions. The supercell size for the interface with steps is 5.33 $\times$ 36.96 $\times$ 40.44 \AA$^3$ and a vacuum gap of 14.1 \AA~thickness separates the slabs. The back side is flattened at the atomic level and the dangling bonds at the top surface of the SiO$_2$ layer and the bottom surface of the SiC substrate are terminated by H atoms. Integration over the Brillouin zone is performed using a 6 $\times$ 1 $\times$ 1 k-point grid including $\Gamma$ point. The grid spacing in real space is taken to be 0.18 $\times$ 0.19 $\times$ 0.18 \AA$^3$. The exchange-correlation functional is treated within the local density approximation.\cite{CanJPhys_58_001200} The projector-augmented wave method\cite{PhysRevB_50_017953} is adopted to describe the electron-ion interaction. We implement structural optimization until all the force components decrease to below 0.05 eV/\AA, whereas the atomic coordinates of the SiC bilayer in the bottom layer and the H atoms terminating C dangling bonds are fixed during the structural optimization. 

The LDOS is calculated as
 \[
 \rho(z,E)=\Sigma_{i,k}\int{|\Psi_{i,k} (x,y,z)|^2 dxdy\times Ne^{-\alpha(E-\epsilon_{i,k}^2)}},
 \]
where $\epsilon_{i,k}$ is the eigenvalue of the wavefunction $\Psi_{i,k}$ with indexes $i$ and $k$ denoting the eigenstate and the k-point, respectively, $z$ is the coordinate of the plane where the LDOS is plotted, and $N(=\sqrt{\frac{\alpha}{\pi}} \cdot \frac{1}{N_k})$ is the normalization factor with $\alpha$ as the smearing factor and $N_k$ as the number of k-points in the Brillouin zone. Here, $\alpha$ is set to 16.4 eV$^{-2}$.

\begin{table}
\caption{Position of topmost interfacial SiC bilayer where CBE states lie. \label{tbl:1}}
\centering
\begin{tabular}{ccc}
\hline \hline 
Interface&Model&Interfacial SiC bilayer \\ \hline
Flat & k1 & 2 \\
     & k3 & 2 \\
     & h1 & 3 \\
     & h3 & 1 \\ \hline
Step (before annealing) & k1/h3 & 2 \\
     & k3/h1 & 4 \\
     & h1/k3 & 3 \\
     & h3/k1 & 3 \\ \hline
Step (after annealing) & k1/h3 & 4 \\
     & k3/h1 & 4 \\
     & h1/k3 & 5 \\
     & h3/k1 & 5 \\
\hline \hline 
\end{tabular}
\end{table}

\begin{figure*}[htb]
\begin{center}
\includegraphics{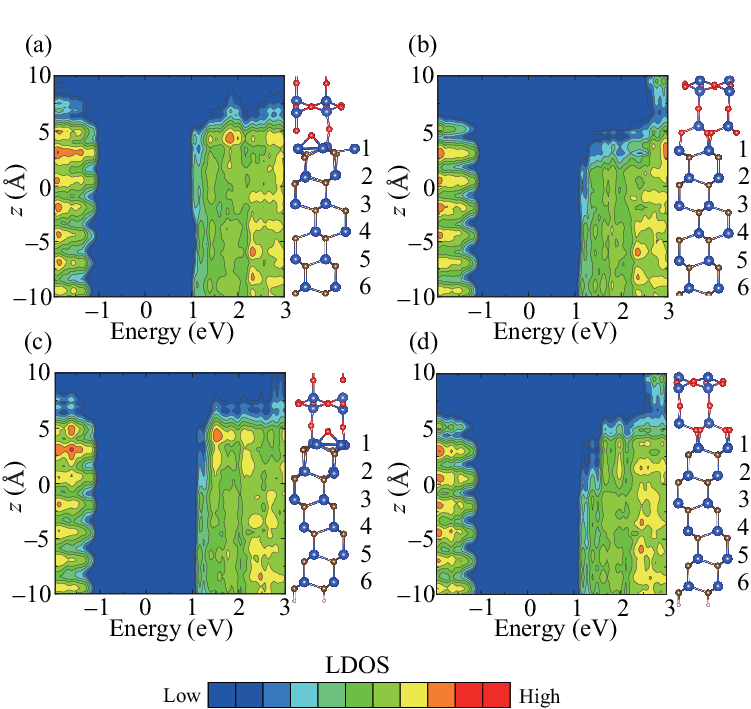}
\caption{LDOSs of interface with steps before annealing for (a) k1/h3, (b) k3/h1, (c) h1/k3, and (d) h3/k1 models. For clarity, structural models are provided on the top of each distribution. Zero energy is chosen as the Fermi level. Each contour represents twice or half the density of the adjacent contours, and the lowest contour is 6.94 $\times$ 10$^{-4}$ electrons/spin/eV/\AA. LDOSs in the shaded area in Fig.~\ref{fig:1} are calculated. Atomic structures are illustrated as a visual aid and the numbers written on the right-hand-side of the atomic structures are the indices of the atomic layers counted from the upper terrace.}
\label{fig:2}
\end{center}
\end{figure*}

Table~\ref{tbl:1} shows the topmost interfacial SiC bilayer where the CBE states appear. As reported in our previous papers,\cite{JPhysSocJpn_85_024701,PhysRevB_96_115311} the CBE states are absent at the topmost SiC bilayer when the first interfacial SiC bilayer of 4H-SiC(0001)/SiO$_2$ is the $k$ site. On the other hand, when the first interfacial bilayer is the $h$ site, the CBE states are absent at the first and second interfacial bilayers in the h1 model, whereas they lie at the first interfacial bilayer in the h3 model.\cite{JPhysSocJpn_85_024701} Matsushita {\it et al.} reported that the behavior of the electrons in the CBE states is similar to that of free electrons, which are called the floating states. 4H-SiC bulks consist of tetrahedrons surrounded by Si atoms and those by C atoms. The electrostatic potential in the Si tetrahedrons is lower than that in the C tetrahedrons because of the difference in electron negativity between the Si and C atoms. The Si and C tetrahedrons spatially overlap at the $k$ site, whereas they are separated at the $h$ site, resulting in the formation of the floating states in the Si tetrahedrons at the $h$ site.\cite{PhysRevLett_108_246404} Since O atoms bridge the Si atoms of the first interfacial bilayer in the h1 model, the electron negativity of the O atoms increases the Coulomb potential at the first interfacial bilayer, resulting in the increase in the energy of the floating states.\cite{PhysRevB_96_115311} The CBE states do not appear at the second interfacial bilayer because the second interfacial bilayer is the $k$ site in the h1 model; the topmost SiC bilayer where the CBE states appear is the third bilayer.

We show in Fig.~\ref{fig:1} the computational models for the interface with steps. Since the atomic structures of the $k$ and $h$ sites appear the same when they are seen from the [$1\bar{1}00$] direction, only the k1/h3 and k3/h1 models are illustrated. Figure~\ref{fig:2} shows the LDOS of the upper terrace of the interface with steps. The regions where the LDOS is integrated are indicated by the shaded rectangle in Fig.~\ref{fig:1}. The positions of the topmost interfacial SiC bilayers of the CBE states are listed in Table~\ref{tbl:1}. In the k1/h3 (h1/k3) model [Fig.~\ref{fig:2}(a) (Fig.~\ref{fig:2}(c))], the topmost interfacial SiC bilayer at the upper terrace is the same as that in the k1 (h1) model. Interestingly, the CBE states are absent at the second and third (the first) interfacial SiC bilayers in the k3/h1 (h3/k1) model  [Fig.~\ref{fig:2}(b) (Fig.~\ref{fig:2}(d))], whereas they appear at the flat interface. Owing to the finite size effect along the direction parallel to the interface at the step edge, the energy of the floating states increases and the states do not lie at the CBE. 

\begin{figure*}[htb]
\begin{center}
\includegraphics{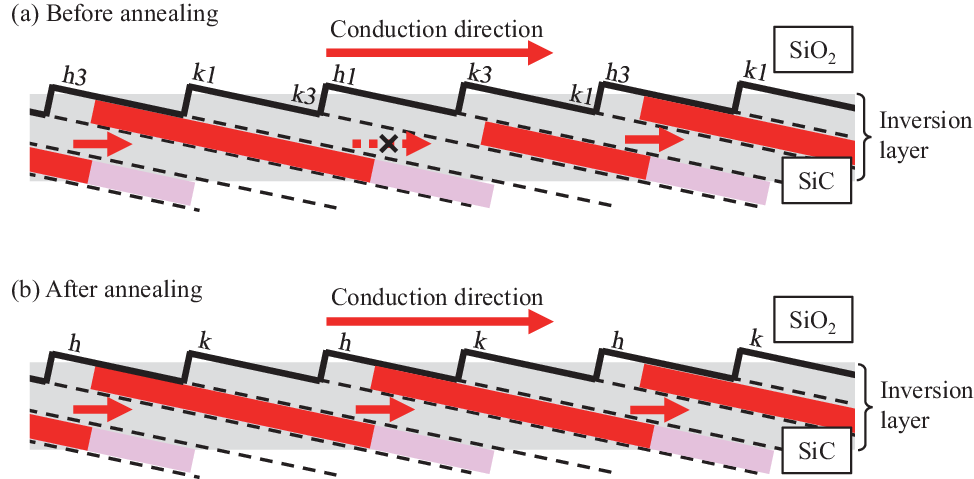}
\caption{Schematic of electron conduction at interfaces (a) before and (b) after annealing. The gray shaded areas indicate the inversion layers. The red and pink areas are the spatial distributions of the floating states below and above the Fermi level, respectively.}
\label{fig:4}
\end{center}
\end{figure*}

The SiO$_2$/SiC(0001) structure is usually fabricated on an off-oriented substrate by 4 degrees and the atomic-scale single steps of the SiC substrate at the interface are observed by STEM. When the gate bias is applied, inversion layers are formed by the CBE states at the interface. Since SiO$_2$ at thermally oxide SiC-MOS is usually amorphous, one-hold and three-hold structures coexist at a terrace and several types of step appear at the interface. The inversion layers formed by the CBE states are not continuous at the step edges in the h1/k3 model. The electrons trapped at the discontinuous inversion layers are immobile, as shown in Fig.~\ref{fig:4}(a), resulting in the low mobile carrier density at the interface.\cite{ApplPhysExpress_10_046601} In addition, since the behavior of the CBE states is sensitively affected by the interface atomic structure, the Hall mobility of SiC-MOS is low.\cite{JApplPhys_131_145701} Therefore, the channel resistance, which is the inverse of the product of the Hall mobility and the mobile carrier density, increases.

\begin{figure}[htbp]
\includegraphics{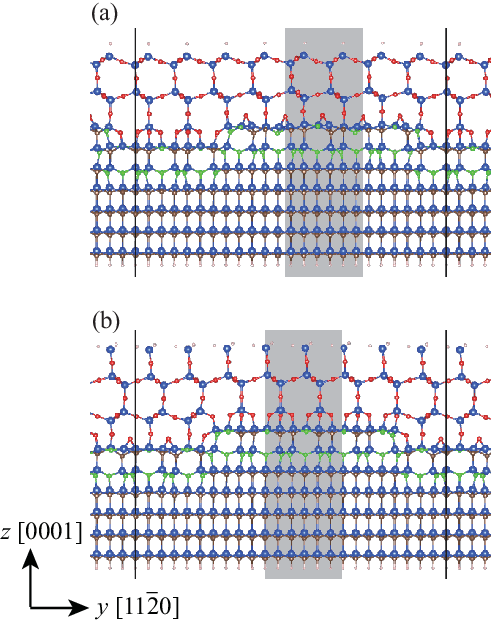}
\caption{Atomic structure of interface with steps after insertion of nitrided layer for (a) k1/h3 (h1/k3) and (b) k3/h1 (h3/k1) models. The meanings of the symbols are the same as those in Fig.~\ref{fig:1}.}
\label{fig:3}
\end{figure}

\begin{figure*}[htb]
\begin{center}
\includegraphics{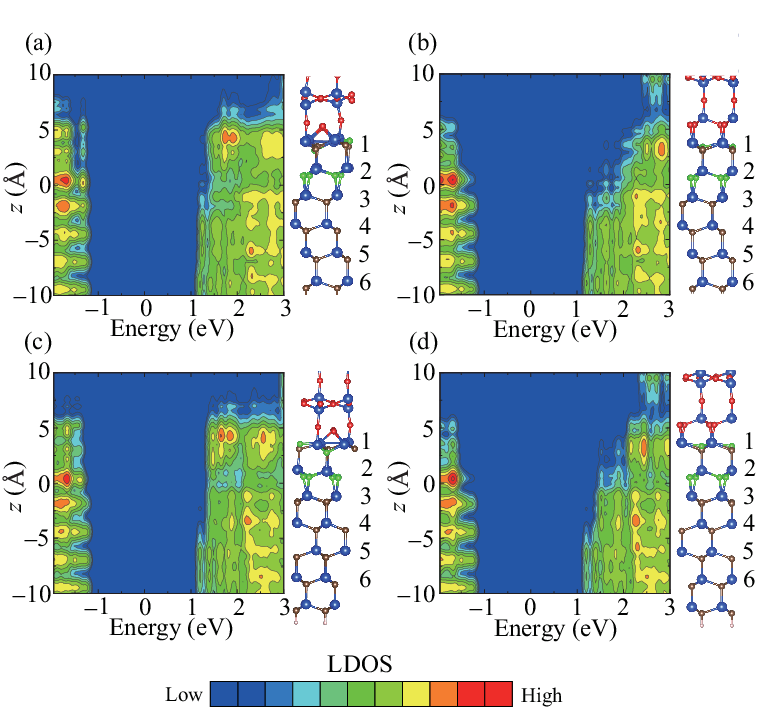}
\caption{LDOSs of interface with steps after annealing for (a) k1/h3, (b) k3/h1, (c) h1/k3, and (d) h3/k1 models. Zero energy is chosen as the Fermi level. Each contour represents twice or half the density of the adjacent contours, and the lowest contour is 6.94 $\times$ 10$^{-4}$ electrons/spin/eV/\AA. LDOSs in the shaded area in Fig.~\ref{fig:3} are calculated. The meanings of the other symbols are the same as those in Fig.~ \ref{fig:2}.}
\label{fig:5}
\end{center}
\end{figure*}

Next, the electronic structure of the interface with steps after NO annealing is investigated. The computational models for the interfaces with steps after NO annealing are shown in Fig.~\ref{fig:3}, in which four C atoms adjacent to a Si vacancy are replaced by N atoms. The structures have been proposed in our previous paper,\cite{JPhysSocJpn_90_124713} and the areal N atom density of 1.2 $\times$ 10$^{-15}$ atoms/cm$^2$ corresponds to that observed in experiments.\cite{JSurfSciNanotechnol_15_109, JApplPhys_97_074902} We have also reported that the insertion of the nitrided layers is an exothermic reaction and the nitrided layers are preferentially formed immidiately below the SiO$_2$ region.\cite{JApplPhys_132_155701} We plot in Fig.~\ref{fig:5} the LDOS of the interface with steps after NO annealing and show the positions of the topmost interfacial SiC bilayers of the CBE states in Table~\ref{tbl:1}. The Coulomb interaction of the O atom in the SiO$_2$ region on the floating states of SiC is screened by inserting the nitrided layer between the SiO$_2$ region and the SiC substrate. The CBE states appear at the fourth (fifth) interfacial SiC bilayer in the k1/h3 and k3/h1 (h1/k3 and h3/k1) models. Because the effect from the O atom in the SiO$_2$ region on the floating states is suppressed, the inversion layers become continuous, as shown in Fig.~\ref{fig:4}(b). The mobile carrier density is increased by removing the discontinuities of the inversion layer at the step edges, where the carriers can hardly penetrate, resulting in the decrease in channel resistance. On the other hand, the nitrided layers do not remove the small deviations of the behavior of the electrons in the CBE states and the Hall mobility remains low, because the insertion of the nitrided layers is not uniform at the practical MOS interface. It is intuitive that the density of these discontinuities is low in the oriented (0001) interfaces. Thus, the investigation of oriented (0001) interfaces without NO annealing is in progress.

We have investigated the effect of NO annealing on the electronic structures of the 4H-SiC(0001)/SiO$_2$ interfaces with steps by the DFT calculations. In the experiments, the mobile carrier density of SiC-MOSs is increased by NO annealing, whereas the increase in Hall mobility is negligible. Our results indicate that the CBE states are absent below the upper terrace in some models before NO annealing owing to the finite-size effect and the Coulomb interaction of the O atom in the SiO$_2$ region, resulting in the discontinuities of the inversion layers under the gate bias. The discontinuities seriously prevent the carriers from penetrating along the channel direction and decrease the mobile carrier density. On the other hand, when the nitrided layers are inserted between the SiC substrate and the SiO$_2$ region by NO annealing, the Coulomb interaction of the O atom is screened and the density of discontinuities in the inversion layers are reduced. Since the spatial deviation of the behavior of the CBE states due to the disorder of the atomic structure at the interface cannot be eliminated completely, the Hall mobility remains low. However, the channel resistance, which is the inverse of the product of the Hall mobility and the mobile carrier density, is reduced. Since the density of the atomic-scale steps, which is the source of the discontinuities of the inversion layers, is low in the oriented (0001) interfaces, these results will also aid future work in determining the channel resistance of SiC-MOSFETs using the oriented (0001) substrates without NO annealing.

\section*{Authors' contributions}
M.U. and N.F. contributed equally to this work.

\begin{acknowledgments}
This work was partially financially supported by MEXT as part of the ``Program for Promoting Researches on the Supercomputer Fugaku'' (Quantum-Theory-Based Multiscale Simulations toward the Development of Next-Generation Energy-Saving Semiconductor Devices, JPMXP1020200205) and also supported as part of the JSPS KAKENHI (JP22H05463), JST CREST (JPMJCR22B4), and JSPS Core-to-Core Program (JPJSCCA20230005). The numerical calculations were carried out using the computer facilities of the Institute for Solid State Physics at The University of Tokyo, the Center for Computational Sciences at University of Tsukuba, and the supercomputer Fugaku provided by the RIKEN Center for Computational Science (Project ID: hp230175).
\end{acknowledgments}

\appendix


\bibliography{main}
\bibliographystyle{apsrev4-1}

\end{document}